\documentstyle[12pt,aasms]{article}

\def\spose#1{\hbox to 0pt{#1\hss}}

\def\kms{\ifmmode {\rm\,km\,s^{-1}}\else
    ${\rm\,km\,s^{-1}}$\fi}
\def\kmsmpc{\ifmmode {\rm\,km\,s^{-1}\,Mpc^{-1}}\else
    ${\rm\,km\,s^{-1}\,Mpc^{-1}}$\fi}

\def\msun{{\rm\,M_\odot}}

\def\ergps{\ifmmode {\rm\,erg\,s^{-1}}\else ${\rm\,erg\,s^{-1}}$\fi}
\def\ergpscm2{\ifmmode {\rm\,erg\,s^{-1}\,cm^{-2}}\else
    ${\rm\,erg\,s^{-1}\,cm^{-2}}$\fi}

\def\deg{\ifmmode {^{\circ}}\else {$^\circ$}\fi}
\def\degr{\ifmmode {^{\circ}}\else {$^\circ$}\fi}
\def\degs{\ifmmode {^{\circ}}\else {$^\circ$}\fi}

\def\h3Mpc{h^{-3}{\rm Mpc}^3}

\def\arcsec{\ifmmode {^{\prime\prime}}\else $^{\prime\prime}$\fi}
\def\asec{\ifmmode {^{\prime\prime}}\else $^{\prime\prime}$\fi}
\def\arcmin{\ifmmode {^{\prime}}\else $^{\prime}$\fi}
\def\amin{\ifmmode {^{\prime}}\else $^{\prime}$\fi}

\def\secper{\ifmmode \rlap.{^{s}}\else $\rlap{.}{^{s}} $\fi}
\def\minper{\ifmmode \rlap.{^{m}}\else $\rlap{.}{^m} $\fi}
\def\arcsper{\ifmmode \rlap.{^{\prime\prime}}\else
    $\rlap.{^{\prime\prime}}$\fi}
\def\arcmper{\ifmmode \rlap.{^{\prime}}\else
    $\rlap.{^{\prime}}$\fi}
\def\spose#1{\hbox to 0pt{#1\hss}}
\def\simlt{\mathrel{\spose{\lower 3pt\hbox{$\mathchar"218$}}
     \raise 2.0pt\hbox{$\mathchar"13C$}}}
\def\simgt{\mathrel{\spose{\lower 3pt\hbox{$\mathchar"218$}}
     \raise 2.0pt\hbox{$\mathchar"13E$}}}

\def\apjref#1;#2;#3;#4 {\par\pp#1, {#2}, #3, #4 \par}

\received{December 22, 1994}

\accepted{March 3, 1995}

\begin{document}

\title{Evolution of IR--Selected Galaxies in $z\sim0.4$ Clusters}

\author{S.\ A.\ Stanford\altaffilmark{1}$^,$\altaffilmark{2},}
\altaffiltext{1}{Astronomy Department, University of California at Berkeley}
\altaffiltext{2}{Visiting Astronomer, Kitt Peak National Observatory, 
National Optical Astronomy Observatories, which is operated by the 
Association of Universities for Research in Astronomy, Inc., under
cooperative agreement with the National Science Foundation.}

\affil{Jet Propulsion Laboratory, California Institute of
Technology, Pasadena, CA  91109 \\
Electronic mail: sas@ipac.caltech.edu}

\author{P.\ R.\ M.\ Eisenhardt\altaffilmark{2},}

\affil{Jet Propulsion Laboratory, California Institute of Technology, 
Pasadena, CA 91109 \\
Electronic mail: prme@kromos.jpl.nasa.gov}

\and 

\author{Mark Dickinson\altaffilmark{1}$^,$\altaffilmark{2}}

\affil{Space Telescope Science Institute, 3700 San Martin Drive, 
Baltimore, MD 21218 \\
Electronic mail: med@stsci.edu}

\begin{abstract}

Wide-field optical and near--IR ($JHK$) imaging is presented for two
rich galaxy clusters: Abell~370 at $z=0.374$ and Abell~851 (Cl0939+47)
at $z=0.407$.  The new data are combined to produce colors sampling
the 0.55--1.65 $\mu$m range in the rest frame. Galaxy catalogs
selected from the near--IR images are 90\% complete to a limiting
magnitude approximately 1.5 mag below $K^\ast$.  The resulting samples 
contain $\sim$100 probable member galaxies per cluster in the central
$\sim$2 Mpc.  Comparison with $HST$ WFPC images yields subsamples of
$\sim$70 galaxies in each cluster with morphological types.  Analysis
of the complete samples and the $HST$ subsamples gives the following
results:

\noindent 1) The optical$-K$ colors of the cluster galaxies are bounded by a 
red envelope of E/S0s which follow a color--magnitude relation with a
slope similar to the corresponding color--magnitude relation in present
epoch E/S0s.

\noindent 2) The $z\sim 0.4$ E/S0s are bluer than those in the 
Bower et al.\ (1992) Coma sample in the optical$-K$ color by $0.13$~mag
for Abell~370 and by $0.18$~mag for Abell~851.  Our estimate of the
systematic uncertainty indicates this color difference is significant
at the 2--3 $\sigma$ level.  If real, the bluing of the E/S0
populations at moderate redshift is consistent with that calculated
from the Bruzual and Charlot (1993) models of passive elliptical galaxy
evolution.

\noindent 3) In both clusters the intrinsic scatter of the known E/S0s about 
their optical$-K$ color-magnitude relation is small ($\sim 0.06$ mag)
and not significantly different from that of Coma E/S0s as given by
Bower et al.\ (1992), indicating that the galaxies within each cluster
formed at the same time at an early epoch.  These results support the
paradigm in which the stellar population in cluster E/S0s is a function
only of the galaxy mass and the lookback time.

\noindent 4) Although the passive evolution models fit the $z\sim 0.4$ 
early--type galaxies in the optical$-K$ color, they fail to reproduce
the observations at intervening wavelengths.  When normalized at rest
frame $\sim$1.6 $\mu$m, the $z\sim 0.4$ E/SO's are fainter than the
model at rest frame $\sim$0.9 $\mu$m, and brighter than the model (and
than Coma E/SO's) at rest frame $\sim$1.2 $ \mu$m.  Such color
differences are difficult to explain in terms of age or conventional
metallicity effects.

\noindent 5) The disk galaxies in Abell 851 are bluer and represent 
a greater fraction of the total sample compared to those in Abell 370.
X--ray maps and large--field optical images indicate that Abell 370 is
``older'' than Abell 851, in the sense of showing a more relaxed
distribution of galaxies and hot gas.  In agreement with the apparent
difference in cluster age, our results suggest the disk galaxies of
Abell 370 have had more time in the cluster environment, relative to
the disks in Abell 851, to redden and/or fade into obscurity.
\end{abstract}

\keywords{galaxy clusters}

\section{Introduction}

The centers of rich galaxy clusters today are dominated by red,
early--type galaxies.  The standard picture for the formation of
elliptical galaxies ($e.g.$, Eggen, Lynden-Bell, and Sandage 1962;
White and Rees 1978) has their stars forming rapidly in a single burst
at high redshift and subsequently evolving passively in a generally
redward march to their present colors.  The color-magnitude relation in
ellipticals is attributed to more massive galaxies retaining their star
forming gas from ejection by supernovae winds more effectively,
resulting in higher metallicities and hence redder colors (Arimoto and
Yoshii 1987).  The apparent coevality and old age of present--day
cluster ellipticals is strongly implied by the small scatter in the
colors at a given magnitude, and the similarity of those colors from
cluster to cluster (Bower et al.\ 1992, hereafter B92).  This suggests
the spectral energy distributions (SEDs) of ellipticals are a function
only of their mass and of the time since their formation.  A sample of
clusters spanning a wide range of redshifts might therefore offer an
excellent means of studying galaxy evolution at large lookback time,
each cluster providing a ``snapshot'' of the evolutionary history of
early--type galaxies.  If the range of formation epochs was small
enough, one could even hope to determine the basic cosmological
parameters $H_0$ and $q_0$ by mapping a range in cluster redshifts to
the corresponding time interval derived from changes in the cluster
elliptical SEDs, in essence synchronizing the cosmological and stellar
evolutionary clocks.

This picture of cluster galaxy evolution was revealed to be overly
simplistic with the discovery of a substantial population of blue
galaxies in distant clusters, which are largely absent in similar
clusters in the present epoch (Butcher and Oemler 1978, 1984).
Spectroscopy of these blue cluster galaxies has revealed them to be a
diverse population comprising star forming systems, ``post--starburst''
galaxies and AGN.  In most cases, however, the blue colors appear to
result from star formation.  Recent $HST$ imaging of clusters at
$z$=0.3 to 0.5 (Dressler et al.\ 1994a and 1994b, Couch et al.\ 1994,
Wirth et al.\ 1994) shows that most blue objects are late--type disk
galaxies, which must have disappeared from the cores of rich clusters
by today, or have been transformed into another type of galaxy.  A
variety of mechanisms have been suggested to explain this process,
including ram--pressure stripping, tidal distruption, galaxy
interactions, and mergers.
 
Despite this rapid evolution in the blue population, ground--based
photometry and spectroscopy and $HST$ imaging shows that a red E/S0
population persists in clusters out to large redshifts and still
dominates the cores of rich clusters at least to $z \approx 0.5$.  The
increasing ratio of disk to spheroidal galaxies with cosmic time,
however, suggests an evolutionary connection between these populations
which would contradict the simple ``single starburst \& passive
evolution'' paradigm for elliptical formation.  Furthermore, Dressler
and Gunn (1983, 1992; hereafter DG92) found that even some of the red
galaxies show spectroscopic signs of recent star formation.  These
``E+A'' galaxies appear to have post--starburst spectra resembling
early--type galaxies with a strong A--star population superimposed (but
no signs of current star formation).  These observations suggest
consideration of the antithesis of the ``synchronized clock'' scenario
described above, namely that all elliptical galaxies form by mergers in
a process which is continuing to the present day.  Tracking the
evolution of the elliptical population in distant galaxy clusters can
not only test the passive evolution scenario, but could help unravel
the Butcher--Oemler effect as well.

With this in mind, we have undertaken a program to trace the
photometric evolution of large numbers of cluster ellipticals over the
widest accessible range of redshift, and hence cosmic epoch.  A
substantial number of galaxy clusters are known out to $z \approx 1$
(e.g. Gunn, Hoessel and Oke 1986; Couch et al.\ 1991) and perhaps
beyond (e.g. Dickinson et al.\ 1995), providing sites for studying
elliptical galaxy colors over more than half the age of the universe.

One difficulty with such a program is that the large redshifts of
distant clusters move optical passbands into the blue and
near--ultraviolet region of the cluster rest frame.  Optically selected
samples may therefore incorporate a redshift--dependent bias toward
star--forming galaxies, which are bright in the rest frame near--UV,
possibly distorting conclusions on galaxy evolution.  This bias can be
minimized by selecting cluster galaxies at near infrared wavelengths.
Infrared photometry samples the peak (in $F_\nu$ units) of the spectral
energy distributions for normal galaxies, measuring luminosities and
colors of the old stellar population which dominates the stellar mass.
The effects of ongoing or recent star formation on galaxy selection are
thus minimized, and the spectral similarity of spirals and ellipticals
in the near--IR insures that a relatively uniform and representative
distribution of galaxy types will be identified across a wide range of
redshifts.  Furthermore, observations in the near--IR are relatively
immune to the effects of reddening and extinction.  These qualities
make the near--IR a valuable regime in which to study distant galaxy
clusters.

Near--IR observations of moderate redshift clusters have only become
technically feasible in recent years. Lilly (1987) used a
single-element detector to obtain photometry of 53 galaxies in five
clusters at $z\sim0.45$.  He found the 36 optically red galaxies in his
sample to be on average $\sim$0.1 mag {\it redder} in rest frame $V-H$
than the average elliptical in the Coma cluster.  While Lilly modelled
this surprising result by incorporating an AGB population in a
passively evolving model, more recent spectral synthesis models
incorporating AGB stars (Bruzual and Charlot 1993; hereafter BC)
predict that single age stellar populations should become monotonically
bluer at earlier epochs, and thus fail to explain Lilly's result.

Arag\'on-Salamanca, Ellis, and Sharples (1991; hereafter AES) published
the first array--based infrared study of galaxy clusters.  Based on a
sample of 46 cluster members down to $K=17.5$ in Abell~370, a rich
cluster at $z=0.374$, they found that the $R-K$ colors lie along a
color--magnitude (c--m) relation similar to that observed for
present--day ellipticals, but with significant scatter to both the red
and the blue.  AES interpreted the redward scatter as arising from AGB
stars in a post--starburst phase, supporting the idea that most cluster
galaxies, including the ellipticals, passed through a starburst phase
at some time after formation.

Arag\'on-Salamanca et al.\ (1993; hereafter AECC) observed 10 clusters
with $0.5 < z < 0.9$ at $V,I$, and $K$.  In contrast to Lilly and to
AES, they found that the optical--IR colors of the reddest cluster
galaxies (the ``red envelope,'' O'Connell (1987)) became progressively
bluer towards higher $z$, in accord with predictions from models for
passive evolution of single--age stellar populations.  Furthermore,
AECC found that clusters at the same redshift showed the same amount of
color evolution, supporting the idea of coeval elliptical formation at
high redshift.  The bluing of the red envelope has also been seen by
Rakos and Schombert (1994) in the optical colors of a sample of high
redshift clusters, and in spectroscopic measurements of the 4000\AA\
break amplitude in clusters at $z > 0.6$ by Dressler and Gunn (1990).
These results are all broadly consistent with the ``synchronized
clock'' scenario for elliptical galaxies outlined above.

The apparently conflicting data on the red envelope and differing
theoretical predictions of the effects of AGB populations suggest that
the history of cluster galaxy evolution is by no means settled yet.
The advent of new, large--format infrared arrays has opened a new era
in which the measurements painstakingly collected by Lilly and by
Arag\'on-Salamanca et al.\ can be substantially improved in both
quantity and quality.  The AECC sample of 10 clusters contained only
$\sim$15 color--selected galaxies per cluster.  With larger arrays, we
can now observe $\sim$100 galaxies per cluster with relative ease in a
single observation.  In addition, with the possibility for
morphological classification of high redshift galaxies from {\it HST}
imaging, the field is clearly open to progress.

Infrared array cameras have been used to obtain multi--wavelength
imaging photometry for $\sim$30 clusters in the redshift range $0 < z <
0.7$.  This dataset consists of $JHK$ imaging reaching a limit $\sim$2
mag below $L^\ast$ over a $\sim$1.5 Mpc diameter field, as well as
complementary optical photometry.  This paper presents our methods of
data reduction, object detection, sample definition, and photometry.
The color evolution of the red envelope is investigated in the first
two clusters observed in the survey, Abell~370 ($z = 0.374$) and
Abell~851 ($z = 0.407$, also known as Cl~0939+47).  Future papers will
present the larger data set for clusters at both lower and higher
redshifts, and investigate the evolution of the galaxy population
across a much broader range of cosmic epoch.  A cosmology with $q_0$ =
0.1 and $H_0$ = 50\kmsmpc is assumed throughout this paper, leading to
a lookback time of 4.5 Gyr for Abell~370 and 4.8 Gyr for Abell~851.
The relative distance modulus between Coma and the two clusters is 6.40
mag for Abell~370 and 6.61 mag for Abell~851 for our adopted
cosmology.\footnote{Adopting $q_0 = 0.5$ changes the relative distance
modulus by approximately --0.15 mag.  This is relatively insignificant
for the analyses presented here (which primarily concern galaxy colors)
because of the shallow slope of the color--magnitude relation.}

\section{Data}

Images of galaxy clusters were obtained over a wide wavelength range to
enable an empirical approach to studying galaxy evolution.  Galaxies
with different redshifts can be compared at constant rest frame
wavelengths (Figure 1).  Furthermore, the relatively unexplored area of
infrared galaxy colors at moderate redshifts can be explored.
 
\subsection{Observations}

Near--IR imaging was obtained of Abell 370 and Abell 851 using SQIID
(Ellis et al.\ 1993) at the KPNO 1.3m telescope.  SQIID uses four PtSi
256$\times$256 arrays fed by a system of dichroic beam--splitters to
provide wide-field ($\sim$5.5 arcmin), simultaneous $JHKL$ band imaging
capability at 1.30 arcsec per pixel resolution.  Data were not saved
from the $L$ band channel because of the high thermal background.
Observations were obtained over the course of three nights in 1991
December, using a two dimensional dither pattern with a 10 arcsecond
step size and 30 arcsecond total extent.  Approximately 115 five minute
frames were obtained in the $JHK$ bands for both clusters.  Several
standard stars from the list of Elias et al.\ (1982) were observed each
night, at 4 separate array positions for each star.  In addition, 5
Coma cluster galaxies from Persson et al.\ (1979) were observed, using
a dither pattern with two arcminute steps and separate off-source sky
frames.  The weather was photometric throughout the time the clusters
were observed.

Optical images of Abell 370 and Abell 851 were obtained through
intermediate bandwidth filters chosen to approximately match the
effective wavelength of the rest frame $V$--band in each cluster (see
Figure 1).  For Abell~370 the relevant optical band is centered at
7520\AA\ ($\Delta \lambda = 375$ \AA), and for Abell~851 at 7840\AA\
($\Delta \lambda = 325$ \AA). Abell 370 was observed at the KPNO 2.1 m
telescope, in photometric weather on 1991 November 2, with the T2KA
chip giving a scale of 0.3 arcsec pixel$^{-1}$.  Abell 851 was observed
also at the 2.1 m, in photometric weather on 1992 April 4, with the
T1KA chip giving a scale of 0.3 arcsec pixel$^{-1}$.
Spectrophotometric standard stars were observed to calibrate the
imaging.

\subsection{Reductions}

The IR data were reduced using DIMSUM\footnote{Deep Infrared Mosaicing
Software, a package of IRAF scripts, available on request from the
authors.}.  The data were first linearized, and then trimmed to exclude
masked columns and rows on the edges of the arrays.  Due to the highly
uniform response of the PtSi arrays, flat--fielding was not performed.
Tests of standard star images flattened with dome flats, sky flats, and
no flats showed no improvement in the repeatibility of measured
standard star magnitudes from flatfielding.  Sky and dark subtraction
were done by subtracting a scaled median of 9 temporally adjacent
exposures from each frame.  Due to the large number of objects in each
field, a first pass reduction was used to produce an object mask for
each frame.  This was done on the fully stacked mosaic image, and thus
the resulting mask excludes objects too faint to be detected on
individual exposures as well as bright objects.  The data were then
rereduced using the object mask to suppress object contamination of the
sky in the production of sky frames.  Final mosaiced images of each
cluster were made with a replication of each pixel by a factor of 4 in
both dimensions, eliminating the need for interpolation when the
individual frames are co--aligned, while conserving flux.  Bad pixels
were excluded from the summed images.  The resulting FWHM of a stellar
image in the summed $K$ band image is $\sim$2.3 arcsec.

Reduction of the optical data was done using standard procedures.  The
optical images were geometrically transformed to match the pixel scale
(0.326 arcsec after pixel replication) and orientation of the IR
images.  A $7520,J,K$ color composite image of Abell~370 is shown in
Figure 4, and a $7840,J,K$ color composite image of Abell~851 is shown
in Figure 5.  Henceforth, the $7520$ and $7840$ filters are referred to
as ``optical''.  Only regions of the final mosaic images with $>$99\%
of the total integration time were used in this investigation, thus
insuring that the sky noise remains very nearly uniform across the
entire field of view.  The resulting fields are $\sim$4.7 arcminutes
square.

\section{Photometry}

The SQIID photometry was transformed to the CIT scale by using
observations obtained each night of several standard stars from the
list of Elias et al.\ (1982).  The standard star solution gave
residuals of $\pm$0.025 mag.  The transformation equations to convert
from instrumental ($jhk$) to calibrated ($JHK$) magnitudes are:

\[ J = j + jzp - 0.14 \times (airmass) \]

\[ H = h + hzp - 0.06 \times (airmass) - 0.21 \times (H-K) \]

\[ K = k + kzp - 0.09 \times (airmass) \]

\noindent
where $jzp, hzp, kzp$ are the instrumental zeropoints.  The $H$ band 
solution includes a large color term in agreement with the one 
determined by Silva and Elston (1994).   

The optical magnitudes were calibrated onto the AB system ($e.g.$ Oke
1979) using spectrophotometric standard stars and then transformed to a
``standard'' (approximately normalized to $\alpha$~Lyrae) system using
$m(7520)_{std} = m(7520)_{AB} - 0.36$, and $m(7840)_{std} =
m(7840)_{AB} - 0.37$.  In our magnitude system, Vega has $m=0$ in the
$JHK$ bands (Elias et al.\ 1982) and $m= +0.03$ in the optical bands,
as it does in the calibration of the $V$ band by Johnson \& Morgan
(1953).

\subsection{Photometric Tests}

An internal check of the $K$ band photometric precision was performed
on a sample of $\sim$100 galaxies in Abell~370.  Photometry in 4.8
arcsec diameter circular apertures was measured on independent datasets
obtained on two different nights.  The differences in the measured
magnitudes for the same objects are plotted against $K$ mag in Figure
2.  Also shown is the expected $\pm 1 \sigma$ errorbar at $K=16.5$
calculated from Poisson statistics, which is only slightly smaller than
the actual scatter.  The calculated rms magnitude differences indicate
that the $K$ photometry in the final, summed images are good to 15\% on
average down to $K=18$.  This is considerably below the estimated
$K^\ast =17.0$, which would be measured in the 4.8 arcsec aperture,
assuming a present--day value $M^\ast(K) =-25.1$ (Mobasher et al.\
1993) and a $k$--correction derived from an unevolving BC elliptical
model.

Two external tests of the IR photometry were performed.  A small number
of Coma cluster galaxies were imaged with SQIID during our 1991
observing run.  The resulting photometry from our data is compared in
Table~1 with that of Persson et al.\ (1979).  For the 5 galaxies in
common with their dataset, the mean differences in magnitudes are:
$J=-0.01\pm0.03, H=-0.01\pm0.02, K=0.00\pm0.02$.  While these galaxies
are $\sim$6 mag brighter than the galaxies in our distant clusters, the
comparison suggests no systematic difficulties with our photometry.
Our $K$ magnitudes for Abell~370 have also been compared with those of
AES.  There are $\sim$43 galaxies in common between the two studies.
The AES photometry was first converted from the UKIRT system to the CIT
system used here (Casali and Hawarden 1992).  Magnitudes were measured
from our data through the same size apertures used by AES.  The
differences are plotted against $K$ in Figure 3.  A systematic offset
is seen in the sense of our measurements being $\sim$0.05 mag fainter
on average than those of AES.  Most of this offset is attributable to
the better seeing reported by AES (2.0 arcseconds compared to our 2.3
arcseconds).  Using model galaxies with appropriate profiles and sizes
(as in \S 4.1 below), our aperture magnitudes should be $\sim$0.03 mag
fainter due to the poorer seeing.  Thus, the comparison shows
reasonable agreement in zeropoint, with a dispersion of 0.15 mag down
to $K=17.5$.

\subsection{The Cluster Catalog}

Object detection was carried out on the sum of the $J$, $H$ and $K$
images in order to minimize false detections at the faint end.  The IR
sum goes significantly deeper than the individual bandpasses.  The
small range of colors for most objects in the IR images ensures that
little or no selection bias is introduced by this procedure, especially
since the final galaxy catalogs are truncated to a limiting magnitude
$K \sim$18 where incompleteness first sets in (see below).

Object detection, cataloging, and photometry was done using the FOCAS
package (Valdes 1982), with some recent modifications and improvements
by Valdes and by the present authors.  To understand its operation on
our data, FOCAS was run on an simulated cluster image having the noise
and object density characteristics of the real $K$ image of Abell~370.
Frames were created to simulate the individual exposures in the raw
dataset.  These raw frames were passed through the reduction procedures
described in \S 2.2.  The advantage of the simulated image is that the
positions and magnitudes of all galaxies are known.  No attempt was
made to include galactic stars, foreground or background field galaxies
in the artificial images.  Experiments with varying the minimum
detection area, the detection level, and the spatial filter were
performed with the FOCAS object detection routine on the artificial
cluster image.  The resulting detections at faint magnitudes were
inspected visually and compared with the known galaxy list to determine
a reliable set of detection parameters.

FOCAS was then used for object detection on the actual cluster data
using the sum of the $JHK$ images as described above.  The detection
parameters were set at the 2.5 $\sigma$ level, with a 9$\times$9
pixel\footnote{When referring to pixels henceforth we will always mean
``sub--pixels'' of the original image, i.e. the 0\farcs326 pixels that
result after the 4$\times$4 replication described in \S 2.1 above.}
($\sim 2\farcs9 \times 2\farcs9)$ tapered boxcar spatial filter, and a
minimum area criterion of 50 pixels, which is about 1.3 $\times$ the
seeing disk area at FWHM.  Sky measurement for each object was made
using $\pm 2 \sigma$ clipping about the mean, a 6~pixel buffer region
around the cataloged area of each object and an 8~pixel wide sky
region.  This ensures that the sky measurement is relatively unbiased
by light at large radii from the object under consideration.  Splitting
of merged objects was done within FOCAS using a reduced minimum area
criterion of 25 pixels to better separate overlapping objects.
Comparison of the results with the optical images (taken in much better
seeing) indicates that the splitting was largely successful in
separating merged objects in the infrared data.  The resulting catalog
was then used to define photometric apertures for objects in all four
IR and optical bands.  In this way, the same isophotal and sky areas
are used for each object in each band.  For the remainder of the
analysis, the catalog was restricted to a limiting aperture magnitude
of $K=18.0$.  The final catalogs contain 162 objects for Abell~370, and
154 for Abell~851.

A test of the FOCAS catalog completeness was made using an artificial
cluster image which simulates the sum of the $JHK$ images of Abell~370.
In each trial, 10 galaxies at a given magnitude were added at random
positions to the artificial image.  Only a small number of galaxies was
added in each trial so as not to greatly increase the galaxy density,
which is an important factor for object detection in a cluster.  FOCAS
was run on this image and the fraction of the extra galaxies found
recorded.  Trials were repeated many times with the extra galaxies
having $K$=16.5, 17.0, 17.5, 18.0, 18.5, and 19.0 to statistically
determine the completeness at these limiting magnitudes. The results
are shown in Figure 6.  The plot indicates that using our adopted
detection parameters produces a catalog of objects that is 90\%
complete at $K=18.0$.

\subsection{Aperture Photometry}

A single--size, metric circular aperture was chosen for the photometry
because such measurements are easy to reproduce and correction for
seeing effects is more straightforward.  Apart from signal--to--noise
considerations, the main disadvantage of this choice is that
single--size apertures measure different fractions of the total light
depending on a galaxy's luminosity, angular size and light profile.
The fact that the galaxies of interest are all at the same redshift in
each cluster, however, ensures that the aperture corresponds to a fixed
metric diameter for all cluster galaxies.  Single--size apertures
should not adversely affect the calculated colors, as long as the
apertures are relatively large, and the effective seeing of the
measured images is the same.  The seeing for the optical images was
$\sim$1 arcsec, much better than for the near--IR data.  Therefore the
optical band images were convolved with a Gaussian to produce an image
with a PSF approximately similar to that of the near--IR images. Galaxy
crowding can also affect simple aperture magnitudes, but for most
objects in these clusters this is relatively unimportant given the
adopted aperture sizes.  The simplicity of the aperture measurements
was judged to outweigh the disadvantages.

Extensive tests of aperture photometry were performed on simulated
galaxies of various sizes produced with the IRAF/artdata tasks to
quantify single--size circular aperture photometry biases on the
measured magnitudes and colors.  These galaxies simulated the range of
apparent magnitude and size expected for elliptical galaxies in the $K$
and optical band images.  Noise characteristics of the Abell~370 images
were imposed on the simulated galaxies, which were produced both with
no seeing and after being convolved with stellar images taken from the
Abell~370 $K$ and optical images.  These tests demonstrated that for an
aperture diameter about twice the seeing FWHM, errors due to differing
PSFs in the optical and $K$ bands are kept below 0.02 mag within the
luminosity range of our sample.  Color gradient information is
preserved in the measured apertures as long as the effective seeing of
the various bands is the same.

Metric aperture diameters of 30 kpc were adopted for the final
photometric measurements, corresponding to 4.5 arcsec for Abell~370 and
4.3 arcsec Abell~851.  Tables 2 and 3 present the resulting catalogs
for Abell~370 and Abell~851, respectively.  Column (1) gives an ID
number from this paper which is ordered consecutively by $K$ magnitude,
and column (2) gives the Butcher--Oemler ID for galaxies in Abell~370
and the Dressler \& Gunn ID for galaxies in Abell~851.  Columns (3) and
(4) give the $x,y$ positions in arcseconds relative to the brightest
galaxy (in the $K$ band) in the cluster, which is marked in Figures 4
and 5.  The J2000 coordinates of the two brightest galaxies are given
in Tables 2 and 3.  Column (5) gives the membership status based on
published spectroscopy (Abell~370: Mellier {\it et al.} 1987;
Abell~851: DG92), where SM and SNM denote spectroscopic members and
nonmembers, respectively.  Column (6) gives the morphological
classification from {\it HST} imaging (described below) with E for
elliptical, L for S0, A for Sa, B for Sb, C for Sc, D for Sd, and M for
merger.  Column (6) also shows an asterisk for those objects determined
to be stars from their shapes in the {\it HST} imaging; however this
determination is not complete---fainter stars are not necessarily
classified as such.  Columns (7--9) give the $KHJ$ aperture magnitudes,
and column (10) gives the optical--$K$ color.  Color--magnitude
diagrams for the catalog objects of the two clusters are shown in
Figure 7.  Errorbars shown at representative $K$ mag indicate the $\pm
1$ $\sigma$ measurement scatter in the colors determined from a
comparison of data obtained on separate nights.

One possible correction to the photometry not discussed yet is galactic
extinction.  Both Abell~370 ($b = -54\deg$) and Abell~851 ($b =
+48\deg$) are at relatively high galactic latitude so low values of
$E(B-V)$ are expected.  From their positions in the maps of Burstein
and Heiles (1982), the predicted values are $<0.03$ for Abell~370 and
$\sim0.0$ for Abell~851.  Values of $E(B-V)$ of 0.01 for Abell~370 and
0.0 for Abell~851 are adopted.  The standard interstellar extinction
curve of Mathis (1990) was used to convert $E(B-V)$ to extinctions in
the relevant bandpasses.  In the photometry reported in Table 2, no
correction has been made for galactic extinction.  The extinction
correction is made later when making comparisons with models and with
the Coma cluster.  An uncertainty of 0.02 mag in the optical--$K$
colors due to the uncertainty in the galactic extinction has been
assumed in calculating systematic errors for both clusters.

\section{Results}

\subsection{Field Correction}
  
The general distribution of objects in Figure 7 for the two clusters is
broadly similar, and resembles that seen in purely optical c--m
diagrams of distant clusters (e.g. Butcher, Oemler and Wells 1983;
DG92).  A sloping ridge line forms a ``red envelope'' which in nearby
clusters is comprised mainly of early--type galaxies.  Not all of the
objects shown in the c--m diagrams are cluster members; field galaxies
contaminate the samples at all magnitudes and colors.  Ideally,
spectroscopic redshifts would be available for all galaxies in the two
fields so that members-only versions of the c--m diagrams could be
constructed.  Compared to other clusters at similar redshifts, these
two have a large number of spectroscopically determined members (48 for
Abell~370, Mellier et al.\ 1987; 26 for Abell~851, DG92).  But compared
to the total number of galaxies in each field, the fraction of
confirmed members is still small.  Also, spectroscopically--selected
samples may be biased towards bluer galaxies. To make better use of the
large number of galaxies in our samples, most of which are expected to
be cluster members, a statistical correction can be made for the field
galaxy population.  Based on the Glazebrook et al.\ (1994) field galaxy
counts in the $K$ band approximately 40 field galaxies should be seen
in our $\sim$5 arcmin square fields to a limiting magnitude of
$K=18.0$.

The field correction was made in the optical$-K$ vs.\ $K$ c--m plane to
preserve the c--m information under investigation.  The details are
described here for Abell~370; the procedure for Abell~851 was similar.
A composite ``field'' c--m diagram was first constructed, using field
galaxy photometry from Glazebrook et al.\ (1994) down to $K=17.25$,
where incompleteness sets in, and from McLeod et al.\ (1994) between
$K=17.25$ and 18.  Glazebrook et al.\ provide $RIK$ for a 552
arcmin$^2$ survey, and McLeod gives $RIK_s$ for a 11.9 arcmin$^2$
survey.  The Glazebrook and McLeod $I-K$ colors were approximately
transformed to the observed optical$-K$ color by computing synthetic
colors on a variety of BC model spectra at various redshifts and
determining the transformation as a function of $R-I$.  A composite
``field'' c--m diagram was constructed from the Glazebrook and McLeod
data with the total number of objects scaled to the solid angle of the
cluster image.  The field c--m diagram was then overlaid on the
Abell~370 optical$-K$ c--m diagram, and for each object in the field
sample, the nearest cluster object in color--magnitude space was
selected and removed.  This method is operationally similar to the
approach used by Dressler et al.\ (1994a) to evaluate field
contamination probability for Abell~851 (see their figure 1), although
they do not actually remove objects from their sample on that basis.
It is important to stress that the objects remaining in our
``field--corrected'' sample are not all expected to be {\it bona fide}
cluster members, but simply have a optical--$K$ vs.\ $K$ distribution
{\it consistent} with that expected for an actual, complete sample of
cluster galaxies.

The field correction resulted in the removal of 38 galaxies from the
Abell~370 catalog.  An additional 2 galaxies were removed because they
are larger and brighter than the two central dominant galaxies yet are
located far from the central regions.  Finally, 13 stars, as determined
from their FWHM sizes in the unblurred optical image or in the WFC
image, were removed to complete the field correction.  This is somewhat
larger than the 9 stars predicted by the $K$ star counts in the
Glazebrook et al.\ survey.  The resulting numbers of galaxies in the
field--corrected samples are 109 for Abell~370 and 99 for Abell~851.
Figure 8 shows the optical--$K$ c--m diagrams for the field--corrected
samples of the two clusters.

\subsection{Photometric Comparison to the Coma Cluster}

The coincidence of our optical and $K$ band data on these $z\sim$ 0.4
clusters with the rest frame $V$ and $H$ bands offers the opportunity
for direct comparison with the photometric properties of present--day
elliptical galaxies.  This approach, which was used by Lilly (1987),
AES, and AECC, avoids the pitfalls of using large $k$--corrections or
relying solely on comparison with the predictions of spectral synthesis
elliptical models.  B92 have published $UVJK$ photometry of a sample of
$\sim$ 50~E/S0s in the Coma cluster ($z$=0.023); unpublished $H$
photometry was kindly provided by R.~Bower.

Before the comparison can be made several corrections and
transformations must be applied to the Coma and moderate--redshift
cluster photometry to place them on the same footing.  The Sandage and
Visvanathan (1978) growth curve was used to transform the 12 kpc
diameter Coma $H$ photometry to the necessary 30 kpc diameter aperture
size.  This aperture correction was typically 0.4 mag.  Luminosity
evolution has not been applied to the Coma photometry.  Any reasonable
amount of luminosity evolution out to $z
\sim 0.4$ would be small and have little effect on the color evolution
being measured.  A seeing correction has been applied to the $K$
magnitudes of the Abell~370 and Abell~851 galaxies so that the
aperture photometry represents the light which would be measured in
the metric sizes without dimming due to seeing.  This correction was
evaluated by comparing aperture photometry on an artifical galaxy
before and after convolution with artificial seeing, and amounts to
$-0.22$ mag.

Because the bandpasses employed here do not precisely match the $V$ and
$H$ bands at the redshift of Coma, a differential $k$--correction is
needed to transform the Coma $V-H$ colors to the redshifted optical$-K$
colors.  The conversion can be done very accurately because of the
close wavelength agreement between the Coma, Abell~370, and Abell~851
bandpasses ($e.g.$ Figure~1).  The actual transformations are obtained
by synthetic photometry using the redshifted standard elliptical model
of BC integrated through the appropriate instrument response functions.

Finally, a correction for color gradients was applied to the Coma
photometry, because the measurements of the moderate redshift cluster
galaxies are made with 30 kpc metric apertures while the B92 photometry
of the Coma sample was made with 12 kpc apertures.  Values for $\delta
(V-K)$ mag per dex in radius between --0.13 and --0.27 have been
published (Peletier 1990; Kormendy and Djorgovski 1989; Schombert et
al.\ 1993; Silva and Elston 1994).  A gradient of $\delta (V-H) =
-0.15$ mag per dex in radius was chosen, assuming the gradient in $H-K$
is very small, resulting in a correction of the Coma $V-H$ colors by
0.06 mag blueward. The color gradient correction at such large radii is
uncertain, however, and recent work even calls into question the
reality of color gradients outside galaxy centers (Reid et al.\ 1994).
If any, the most likely error made in the color gradient correction is
to have made the Coma E/S0s too blue in rest frame $V-H$.

It behooves us at this point to carefully review the sources of error
in the comparison of the Coma E/S0s with the moderate--$z$ E/S0s.
Random measurement errors have already been discussed in Section 2.2,
but systematic errors can be important when trying to ascertain small
absolute color differences.  The sources of systematic error in the
comparison of the optical$-K$ colors with the B92 Coma sample are
summarized in Table 4 with their estimated sizes.  Adding these in
quadrature gives an estimated 0.06 mag of total systematic error.

In Figure 8, the transformed colors from the B92 Coma sample are
plotted in the field-corrected optical--$K$ vs.\ $K$ diagram for the
two distant Abell clusters.  The transformed colors are those which
would be measured if Coma were observed at $z \approx 0.4$ through the
same filters used here.  In both clusters, the Coma E/S0 colors form a
red border to the moderate--redshift galaxies.  The red envelope, which
is probably cluster E/S0s, tends to fall somewhat blueward of the mean
Coma locus.

\subsection{Color Evolution in the E/S0 Populations}

Distant clusters have late--type disk galaxies which may affect color
comparisons with the Coma E/S0 galaxies.  Fortunately, morphological
types are now available for large numbers of galaxies in both of the
distant clusters thanks to high resolution {\it HST} imaging.  Hubble
classifications were kindly provided by A.~Oemler (personal
communication) based on WFPC images for Abell~370 and WFPC2 images for
Abell~851.  Within the 2.5 arcmin field of the {\it HST} images, every
galaxy in our IR samples could be assigned a morphological type, so
that a subsample derived from cross--correlation of the {\it HST} data
with our near--IR sample essentially remains an IR--selected sample.
The resulting IR/HST samples contain 79 objects for Abell~370 and 59
for Abell~851.  Within the central 2.5 arcmin of these fields, most
objects in the two samples will be cluster galaxies.  The field
correction for the IR/HST samples described above would predict 9 field
objects for Abell~370 and 11 for Abell~851 within the {\it HST} field
of view.  Because of this relatively small number of predicted field
objects in the IR/HST samples, and due to the difficulty of properly
determining a field correction with morphological as well as color and
magnitude dimensions, the IR/HST samples will be used without
field--correction.  These subsamples are plotted in Figure 9 for
Abell~370 and Abell~851, again in the optical--$K$ vs.\ $K$ diagram.
To simplify the plots, only a linear fit to the transformed colors of
the Coma E/S0s is shown.  The morphological types are represented by
the various symbols as shown in the plots.  The difference between
neighboring types in the Hubble sequence is not highly significant,
particularly for Abell~370 where classifications were determined from
aberrated {\it HST} images (see also Dressler et al.\ 1994b).
Spectroscopic membership information is included in Figure 9, showing
that most of the galaxies are indeed cluster members in the limited
areas of the IR/HST samples.

Returning to the color evolution test, in both clusters the E/S0s tend
to fall to the blue of the Coma E/S0 line.  The median difference in
the optical$-K$ between the Coma E/S0 galaxies and the Abell 370 E/S0s
is $0.13 \pm 0.01$ mag, and the same quantity for the Abell 851 E/S0s
is $0.18 \pm 0.01$ mag.  The given uncertainties are not dispersions;
they are on the medians.  These color differences would be
similar---0.01 mag smaller---if calculated using field-corrected
versions of the IR/HST samples.  Statistical tests were performed on
the optical$-K$ colors of the Coma vs.\ moderate-$z$ E/S0 samples.  The
Kolmogorov--Smirnov test shows that the probability that the Abell~370
E/S0s are drawn from the same parent population as the Coma E/S0s is
4\%, while the probability for Abell~851 is 0.01\%.  The K--S test
suggests that there are formally significant color differences between
the Coma E/S0 galaxies and those in Abell~370 and in Abell~851.
However, systematic error is ignored by the K--S test.  After taking
into account the systematic error, the color differences for E/S0
galaxies in the two distant clusters relative to Coma are significant
at only the 2$\sigma$ level for Abell~370, and at the 3 $\sigma$ level
for Abell~851.

The standard deviation of the optical$-K$ color residuals (the colors
of the cluster galaxies minus the corresponding value of the Coma E/S0
fit at the same magnitude) was calculated to be 0.15 mag for both
clusters over the entire IR/HST E/S0 samples.  The rms scatter in
optical$-K$ colors expected from random measurement errors alone is
estimated to be 0.11 mag, based on an average over the appropriate
magnitude range in Figure 2 (errors in the $K$ magnitude dominate the
observational scatter in the optical$-K$ color).  By using the color
residuals inflation of the scatter due to the c--m relation is avoided.
Because the ellipticals tend to be relatively bright, a better
estimator of the scatter is obtained by only considering galaxies with
$K < 17$.  Down to this limit the rms scatter due to measurement error
is 0.06 mag in the optical$-K$ color, while the standard deviation of
the optical$-K$ color residuals for the E/S0s is 0.085 for Abell~370
and 0.086 for Abell~851.  Thus, the component of intrinsic scatter in
the optical$-K$ color of the moderate$-z$ E/S0s brighter than $L^\ast$
is 0.06 mag for both clusters.  The relatively small intrinsic scatter
for the two moderate$-z$ clusters is similar to the 0.05 component of
intrinsic scatter found by B92 in $V-K$ for their Coma E/S0 sample.

\subsection{Morphology and Color}

The morphological classes in the c--m diagrams have similar color
distributions within each cluster.  The median optical$-K$ color
residuals for each Hubble type are given in Table 5, which demonstrates
the remarkable weakness of the color--morphology dependence for
galaxies ranging from ellipticals to types as late as Sc.  Note that
the $magnitude$ of the bluing of the late--type galaxies in Table 5
does not necessarily represent an appropriate measure of passive
evolution.  To that end, a comparison would have to be made with a
sample of $late$--type galaxies from the Coma cluster.  The Coma
zeropointing is employed to remove the c--m relation, and to enable
meaningful color comparisons between the two $z\sim$ 0.4 clusters.  The
weak trend with morphological type seen in Table 5 is similar to the
results given by Aaronson (1978) for a sample of nearby field galaxies,
and by Bershady (1994) for a $z=0.1 - 0.3$ field sample.  Apparently,
even the fairly substantial star--forming activity found in Sc galaxies
has little effect on their spectra longward of $\sim$0.5 $\mu$m.
Unfortunately, no published optical$-K$ colors exist on a suitable
sample in a present--epoch cluster for the more direct comparison of
our data.  Aaronson (1978) observes that the Sc galaxies in his nearby
field sample are $\sim$0.3 mag bluer in $V-K$ than the E/S0s.  The
color difference between late--types and early--types in nearby
clusters is even less in an optical color (Oemler 1992), and probably
in an optical--IR color too.  The bluing trend with galaxy type is
nearly absent for Abell 370, while in Abell 851 the amplitude of the
trend is similar to that seen in field galaxy populations.  As for the
measured dispersion, the E/S0s are similar in the two clusters, but the
late--type galaxies show wider variation in Abell~851.  The result is
that the overall scatter (averaged over the entire IR/HST samples) is
slightly larger in Abell~851 (0.21 mag) than for Abell~370 (0.18 mag).
The implications of the differences and similarities of the color
distributions among the Hubble types and between the two clusters will
be taken up below.

\subsection{Theoretical Models}

The passively--evolving elliptical spectral synthesis model of BC
provides a moderately good fit to the broadband $UVJHK$ colors of an
$L^\ast$ present--epoch elliptical in the Coma cluster (B92).  So it is
of interest to compare the BC models to the SED of an average $L^\ast$
E/S0 in the two distant cluster samples.  To make the comparison shown
in Figure 10, the slope of the c--m relation was used to correct the
colors of all morphologically classified E/S0s with $L > L^\ast$ (where
random errors are small) to their equivalent colors at $L^\ast$.  These
colors were averaged to produce a representative early--type SED for
each cluster, which was then adjusted for the effect of color gradients
from the measured aperture of 30 kpc to 10 kpc, appropriate for
comparison to the BC model.  The results are plotted in the rest frame
in Figure 10, normalized to the observed $K$--band flux point (i.e.
approximately at rest frame 1.6 $\mu$m), and with 1$\sigma$ scatter
bars shown.  Similar average E/S0 colors for the B92 Coma sample are
also plotted for comparison. The solid line spectrum shows the
no--evolution BC elliptical model.  The dashed line shows the BC
passive--evolution elliptical model for Abell 370, a single--burst
model with an age of 9.0 Gyr.  The passive evolution model for Abell
851, which would be similar at all rest wavelengths to the Abell 370
passive evolution model, was omitted from the plot for clarity.
 
The bluer optical--IR color of the passive evolution BC model compared
to the Coma colors provide a better match to the observed data for
Abell 851 than for Abell 370.  The rest frame $\sim$0.9 and $\sim$1.2
$\mu$m fluxes of the distant E/S0s are not well fitted by the
no--evolution model, and are significantly brighter at rest frame
$\sim$1.2 $\mu$m than the Coma E/S0s.  The Abell 851 E/S0s are fitted
by the passive--evolution model at rest frame $\sim$1.2 $\mu$m.  Both
$z\sim$ 0.4 clusters are fainter than both models at rest frame
$\sim$0.9 $\mu$m.  It is worth recalling that SQIID obtains {\it
simultaneous} $J$, $H$ and $K$ measurements, thus minimizing possible
sources of systematic uncertainty in color measurements, $e.g.$, those
from differences in atmospheric transparency or seeing.  The same does
not apply to the optical--IR colors.

The infrared color mismatch is seen throughout the luminosity range
spanned by our data.  Figure 11 compares the $H-K$ c--m diagrams for
the IR/HST samples to the average relation for Coma E/S0s.  A similar
diagram cannot be made for the $J-K$ color because of a lack of Coma
data at the rest frame $\sim$ 0.9 $\mu$m sampled by the observed $J$
band in the moderate redshift clusters.  Instead, Figure 12 shows the
$J-K$ c--m diagrams with the BC no--evolution and passive--evolution
model colors plotted.  No c--m information is involved in the model
colors in Figure 12 which are most appropriate at $K^\ast \sim 17$.
However, the near--IR c--m relation is shallow so the comparison of the
plotted model colors over the entire magnitude range of the samples is
reasonable.

\section{Discussion}

\subsection{The Early--Type Galaxies}

\subsubsection{Color-Magnitude Relationship}
The well-known c--m relationship seen in nearby galaxy clusters is also
apparent in the optical$-K$ colors of the galaxies in our two moderate
redshift clusters.  The slopes for the E/S0 galaxies in the optical$-K$
color are $0.05\pm0.02$ and $0.07\pm0.01$ mag per mag for Abell 851 and
Abell 370, respectively.  These slopes are determined from linear fits
to the colors of the IR/HST samples, with one bright, blue galaxy
excluded in each cluster because of their unwarranted effect on the
weighting of the fitting.  In Abell 851 the excluded galaxy is a known
spectroscopic nonmember.  Both slopes are similar to the $0.07\pm0.01$
mag per mag of the transformed Coma galaxies.  The lack of a slope
change between the present epoch and $z\sim0.4$ is not too surprising.
Spectral synthesis models which incorporate metallicity suggest that
the slope, which is thought to be due to metallicity variation with
luminosity, will change perceptibly only at lookback times exceeding
7--8 Gyr (Arimoto and Yoshii 1987).  So testing such a model prediction
must await analysis of higher redshift clusters in our sample.

\subsubsection{Optical--IR Color Evolution}
The c--m diagrams shown in Figures 8 indicate that the reddest galaxies
in these two clusters at $z \approx 0.4$ have slightly bluer
optical$-K$ colors than do early--type galaxies in the Coma cluster.
Morphological classifications from {\it HST} confirm this result for
subsamples of galaxies specifically chosen to have E/S0 Hubble types.
Assuming the disk galaxies in Abell 370 and Abell 851 have current star
formation typical of spirals in the present epoch, the relatively weak
dependence of optical$-K$ color on Hubble type (Table 5) strongly
suggests that the observed bluing in the moderate redshift E/S0s is
{\it not} primarily the result of ongoing or recent star formation, but
instead reflects the passive evolution of the old stellar population
which dominates the light in the rest frame 0.55--1.6 $\mu$m spectral
range.  Moreover, none of the E/S0s in Abell~851 (DG92), and only 2 (of
52) of the E/S0 galaxies in Abell~370 show signs of recent star
formation ($e.g.$, strong Balmer line absorption) in optical spectra
(Couch et al.\ 1994).

The color difference between galaxies in the distant clusters and Coma
is small, and in the case of Abell~370 may not be significant given the
amplitude of our systematic errors.  Even if the bluing trends are not
strictly significant with respect to the present epoch, it is clear
that the moderate redshift E/S0s are {\it not} redder, in disagreement
with observations by Lilly (1987) and AES.  The contradiction probably
results from increased data quality afforded by modern detectors,
increased sample size, and better present epoch comparison data of B92.
Our results also do not agree with those presented by Rakos and
Schombert (1994) showing evidence of an AGB effect at $z\sim$ 0.4 based
on a comparison of the optical colors of a sample of high redshift
clusters with the models of Guiderdoni and Rocca-Volmerange (1987).
The redder optical colors of the AGB effect should be even more evident
in our optical$-K$ colors; however, no such effect is predicted by the
BC models.  The slightly bluer optical$-K$ colors of the $z \sim 0.4$
clusters are fairly well matched by simple models of passive spectral
evolution, and agree with the trend seen by AECC for the bluing of the
red envelope in clusters at $z > 0.5$.

Before comparing the optical$-K$ colors between the two $z\sim$ 0.4
clusters, it is of interest to compare the global properties of Abell
370 and Abell 851.  The optical and x--ray morphologies of the two
clusters exhibit striking differences.  The galaxy distribution of
Abell~370 is centrally concentrated and relatively symmetric.  Its two
dominant galaxies are reminiscent of the configuration seen in the Coma
cluster.  The x--ray morphology as observed by the Rosat HRI (Mellier
and Bohringer 1994) is bimodal with peaks associated with each of the
two dominant galaxies, but is otherwise fairly round and centrally
peaked.  The presence of the famous giant luminous arc, faintly present
in our IR images, strongly argues for a sharply concentrated central
mass distribution (Grossman and Narayan 1989; Wu and Hammer 1993).  The
galaxy distribution of Abell~851, by contrast, is more clumpy and
irregular.  Wide--field optical images show several apparently rich
subclusters situated at large radii, plausibly still infalling toward
the central cluster (Dressler et al.\ 1994a; D.\ Silva, personal
communication).  Rosat PSPC observations by Porter (1994) tell the
story most strikingly.  While quite x--ray luminous, the cluster is
large and highly irregular, with several clear sub--clumps visible, and
a low average surface brightness lacking the strong central
concentration seen $e.g.$ in Coma or Abell~370.  The visual impression,
evident in our images, that Abell 370 is richer than Abell 851 is
somewhat misleading.  A good case can be made that Abell~851 is
dynamically younger than Abell~370, and is being viewed at an earlier
stage in its formation.

Bearing the foregoing discussion in mind, it is of interest that the
optical$-K$ colors of early--type galaxies in Abell~851 are 0.05 mag
bluer than those in Abell~370.  The passive evolution model of BC
predicts a difference of 0.02 mag in the optical$-K$ color due to the
difference in cosmic time between the two redshifts (only $\sim$0.3~Gyr
for $H_0$=50\kmsmpc).  The significance of the resulting 0.03 mag
difference is tempered by the systematic error, although many of the
sources of uncertainty given in Table~4 cancel out for a comparison
only at $z\sim$ 0.4.  If real, the optical$-K$ color difference could
be due to younger ages for the early--type galaxies in Abell~851. Taken
at face value, this could imply non--coeval histories for the
elliptical galaxy populations in the three clusters, either due to
different formation redshifts or to important differences in their
recent star formation histories.  Here, ``recent'' means long enough
ago so that evidence of star formation is no longer seen in the optical
spectra of the cluster E/S0s.  Given the small amplitude of the
observed effect, the suggested non--coeval formation cannot be
considered a firm result from the present data.
 
\subsubsection{The History of the Stellar Populations}
The intrinsic scatter found in the optical$-K$ colors of E/S0s in the
two moderate redshift clusters offers a means of placing limits on the
galaxies' stellar population history.  Considering the scatter at a
given magnitude largely avoids the complications due to the
age--metallicity degeneracy.  The scatter in the colors could result
from several factors, including the age of the formation epoch, the
duration of the formation epoch, and recent starburst activity.  The
latter factor is particularly of interest with respect to understanding
the Butcher-Oemler effect.

Along with the Bruzual (1983) models, B92 used the intrinsic component
of the scatter in the Coma E/S0 $U-V$ colors to place limits on the age
and coevality of the formation epoch.  They found that the Coma E/S0s
must have formed at least 13 Gyr ago (for q$_o$=0.1) if the formation
process lasted on the order of the galaxy collapse timescale of $\sim$1
gyr.  These results are in rough agreement with those of Rakos and
Schombert (1994) for the formation epoch of the red population in a
sample of high redshift clusters.  Based on passive evolution model
fits to their optical colors, Rakos and Schombert (1994) put the age of
the early--type population at greater than 15 Gyr.  B92 also explored
limits on the role of recent starbursts, specifically with an eye
towards Butcher--Oemler galaxies at moderate redshifts.  They concluded
that the intrinsic scatter in the Coma E/S0s $U-V$ colors allowed for
no greater than 10\% by mass starbursts in the early--type populations
at $z\sim$ 0.5.

The fact that the intrinsic scatter in the optical$-K$ colors of the
E/S0 populations in Abell 370 and Abell 851 is not significantly
greater than that seen by B92 for a similar color on their Coma E/S0
sample suggests that there has been little starburst activity in the
star formation history of cluster early--type galaxies since $z\sim$
0.4.  This argument assumes that the moderate redshift E/S0s in our
samples evolve into the corresponding Coma population.  But, it is
difficult to use our optical$-K$ colors to place interesting limits on
starbursts at $z > 0.4$ because they are relatively insensitive to the
effects of recent star formation ($e.g.$, Aaronson 1978), unlike the
$U-V$ colors used by B92 for this purpose.  Only very recent
starbursts, on the order of a few $\times$ 10$^8$ years, can be ruled
out as having caused the scatter in the intrinsic optical$-K$ colors of
the E/S0s in Abell 370 and Abell 851.

Our optical$-K$ colors $can$ be used with the BC models to calculate
interesting limits on the epoch and coevality of early--type galaxy
formation under the assumption of passive evolution in the original
stellar populations.  If a Salpeter IMF with mass limits of 0.1 $\msun$
and 125 $\msun$ and a 1 Gyr burst are assumed in computing a BC
elliptical model, then the E/S0s must have formed at least 5 Gyr prior
to the cosmic time at $z\sim$ 0.4 if the spread in the formation time
of the individual ellipticals was $\sim$1 Gyr.  Decreasing the
formation spread time to 0.5 Gyr allows the minimum formation time to
be only 4 Gyr prior to $z\sim$ 0.4.  It is important to note that these
formation epoch limits are $only$ based on the observed scatter, and
that they ignore the absolute colors.  Any age less than about 5 Gyr at
$z\sim$ 0.4 would give unacceptably blue optical$-K$ colors compared to
the observed E/S0 colors in Abell 370 and Abell 851.  Increasing the
formation spread time to 2 Gyr would force the formation epoch to be at
least 7 Gyr prior to the cosmic time at $z\sim$ 0.4.  The BC passive
evolution model which roughly fits the observed optical$-K$ colors of
the IR/HST samples' E/S0s gives a present epoch galaxy age of 13.5 Gyr
with our adopted cosmology.  The lower limit of 10 Gyr to the present
age of the E/S0s in Abell 370 and Abell 851 based on the optical$-K$
scatter approximately agrees with the age of our passive BC model.
Qualitatively, the small scatter in the $z\sim$ 0.4 optical$-K$ colors
agrees with the result of AECC that the early--type galaxies in $z >
0.5$ clusters formed at the same time at high redshift.

\subsubsection{The IR Colors}
The curious departures of the other infrared flux points for the two $z
\sim$ 0.4 clusters from the model caution against
overinterpretation of comparisons with existing evolutionary synthesis
models like those of BC---other effects such as metallicity may play an
important and as yet unquantified role.  The rest frame $\sim$1.2
$\mu$m flux density of the Abell~370 E/S0s deviates further from the
model than that for Abell~851, while the opposite is the case at rest
frame $\sim 0.9\mu$m.  These differences cannot be explained by
standard passive evolution, nor by starbursts.  The B92 sample does not
provide a check on the BC model at rest frame $\sim$0.9 $\mu$m, so one
may only wonder if the model provides a good fit to present epoch
galaxies at these wavelengths.  At rest frame $\sim$1.2 $\mu$m the Coma
E/S0s are actually fainter than the models so that the moderate
redshift E/S0s are even more discrepant relative to our ``empirical''
model of the present epoch.

This spectral region is poorly understood for present epoch galaxies
and could be subject to metallicity effects during stellar evolution in
galaxies.  Some relevant work has been done by Frogel, Terndrup and
collaborators on the colors of Galactic bulge M giants, which probably
dominate the stellar light redward of 1 $\mu$m in elliptical galaxies.
Frogel et al.\ (1990) found the unusual property that the $J-H$ colors
of M giants get bluer with increasing metallicity for stars of similar
temperature.  They attribute the effect to a decrease in the size of
the $H$ band bump in the bulge giants, which may be linked to their
greater metallicity.  Increased H$_2$O absorption in the metal--rich
bulge giants was originally proposed by Frogel and Whitford (1987) as
the explanation.  The greater steam absorption would make up for the
opacity minimum of the H$^-$ ion at 1.6 $\mu$m which causes the $H$
band bump.  However, subsequent work by Terndrup, Frogel, and Whitford
(1991) does not support this conclusion.  Terndrup et al.\ (1990)
suggest that synthesis models of ellipticals should use bulge M giants
instead of the solar neighborhood spectra used in $e.g.$ the BC models.
The effect on the models plotted in Figure 10 probably would be to make
the rest frame 1.2 $\mu$m--1.6 $\mu$m color bluer and thus more closely
match the $z\sim$ 0.4 E/S0s.

However, a bulge--giant BC model would fit the B92 Coma sample worse at
these wavelengths.  The present epoch BC model with local M giant
spectra is already bluer at rest frame 1.2 $\mu$m--1.6 $\mu$m by about
0.05 mag than the B92 E/S0s.  Using the even bluer bulge M giant
spectra would increase the discrepancy at these wavelengths.  It would
also make the rest frame $V-H$ comparison of the BC model with B92
worse in the sense of the model being too blue.  One way out of this
latter problem would be to use an older model.  For example, a BC model
with a present age of 15 Gyr would be redder in rest $V-H$ than the
13.5 Gyr model used in this paper so presumably would fit the B92
colors if bulge M giants, which do not affect the rest $V$ band light,
were used in place of the solar neighborhood M giants.  The effect of
using an older, bulge--M--giant BC model on the optical-IR comparison
with the z$\sim$ 0.4 E/S0s would be only a small change; the bulge
giants act to decrease the IR light while the older age acts to
increase the IR light.  The probable result is that the Abell 370 and
Abell 851 E/S0s would still be fit by the passive evolution BC model.

In the scenario described above, another consideration is the way the
models are brighter at rest frame $\sim$0.9 $\mu$m than our moderate
redshift clusters.  Frogel et al.\ (1990) also find that TiO absorption
around 0.8--0.9 $\mu$m is stronger in bulge M giants compared to local
giants.  So if bulge M giants were substituted for their solar
neighborhood counterparts in the BC stellar spectra library the result
should be a decrease in flux in the BC elliptical model around 0.8--0.9
$\mu$m, which would better fit the $z\sim$ 0.4 E/S0s.

The discussion above is only qualitative but it does suggest that the
unexpected colors of the moderate redshift E/S0s between rest frame
$\sim$0.9 $\mu$m and $\sim$1.2 $\mu$m might be related to metallicity
effects.  Current work on the isochrone synthesis technique includes
expanding the input spectra and stellar evolutionary tracks to include
non--solar metallicities, so that in the future the above speculation
may be tested (Charlot, personal communication).  Perhaps the most
worrisome issue is the implied metallicity differences between the Coma
and moderate redshift E/S0s.  Terndrup et al.\ (1990) find that
$\Delta$[Fe/H]=0.3 between the solar neighborhood and bulge M giants.
The implication is that the metallicity of the $z\sim$ 0.4 E/S0s would
have to change by this amount if they are to become their Coma
equivalents by the present epoch.  An interesting observation to obtain
with regard to this scenario would be rest frame $K$ band imaging of
the $z\sim0.4$ clusters.  Besides the variation of the $J-H$ color with
metallicity, Frogel et al.\ (1990) find that $H-K$ gets redder with
increasing metallicity for M giants.  Though technically difficult,
observations of moderate redshift clusters in the $L$ band, which
samples the rest frame $K$, is becoming feasible for $L > L^\ast$
galaxies.

\subsection{The Late--Type Galaxies and the Butcher--Oemler Effect}

The optical$-K$ colors of galaxies in Abell~851 have a somewhat larger
scatter than do those in Abell~370 (see Figures 8 and 9).  Inspecting
the distribution by morphological type in the IR/HST sample, it appears
that much of this difference is attributable to a population of
late--type galaxies in Abell~851 that are both relatively more numerous
and bluer than those in Abell~370.  Within the {\it HST} field of view,
spirals comprise 46\% of the IR/HST sample of galaxies in Abell~851 and
34\% of those in Abell~370.  Late--type (Sc/Sd) spirals amount to 12\%
of the Abell~851 population compared to only 4\% for Abell~370.
Moreover, the Abell~851 disk systems are significantly bluer ($\sim
0.2$ mag for all spiral types) than those in Abell~370 (Table 5).
Although not directly quantifiable in terms of the classical
Butcher--Oemler fraction, the differences in the spiral populations
suggest a stronger Butcher--Oemler effect in Abell 851.  This result is
in rough agreement with the classical B--O fractions for the two
clusters, $f_B\sim$ 0.2 for Abell 370 and $f_B\sim$ 0.3 for Abell 851
(Butcher and Oemler 1984; DG92).

The tendency for spirals in present--day rich clusters to be redder
than those in the field or in poorer groups has been noted previously
($e.g.$, Oemler 1992).  On average, spirals in Abell~370 have
optical$-K$ colors little different from those of the early--type
galaxies.  Similar results are seen for the average $B-V$ colors of
nearby cluster galaxies tabluated by Oemler (1992).  But at $z\sim$ 0.4
the richer environment, Abell 851, has the bluer spiral colors.  This
difference from the present epoch could correlate either with total
system richness or with the dynamical maturity of the clusters.
Comparisons with dynamically immature rich clusters and with
dynamically mature poor clusters at moderate--redshifts might solve the
ambiguity.  Photometric data in the same optical--IR system on field
spirals at $z \sim 0.4$ also would be valuable for judging whether it
is the Abell~851 disks that are bluer than is to be expected for a rich
cluster at moderate redshifts, or the Abell~370 spirals that are too
red.  From the perspective of present epoch clusters, the Abell~851
spirals are too blue.  Dressler et al.\ (1994a) find that most galaxies
in Abell~851 have rest frame $B-V$ colors consistent with those of
present--epoch field galaxies with comparable Hubble types, with the
exception of a tail of faint Sd/Irr galaxies bluer than their
present--day field counterparts.  (The latter are too faint to be in
our IR/HST samples.)  Moreover, they see little evidence that
interactions and mergers induce bluer colors in the galaxies involved.
These results suggest that the Abell~851 spirals are too blue for a
rich cluster.

Taken together, these points suggest a ``thumbnail sketch'' in which
the presence of a bluer, more numerous spiral population in Abell~851
correlates mainly with a younger dynamical age.  The disk galaxies,
which have colors more similar to field disks than to cluster disks,
may be falling into Abell 851 from the field or from within loose
groups for the first time.  In this scenario, the processes which
transform the blue, disk galaxy population in rich clusters would have
had more time to operate in Abell~370.  Star formation has slowed in
the Abell~370 disk galaxies, leaving them to evolve toward redder
colors.  It may even be that this mechanism disrupts or removes disk
galaxies, reducing their specific frequency.  The smaller spiral
fraction in Abell~370 is consistent with requiring such a mechanism,
but statistics based on only two clusters (so far) cannot be considered
as conclusive.  Infrared selection is particularly valuable for
investigating this sort of question, since it minimizes the possibility
of bias toward star forming objects when tallying galaxies as a
function of morphology or color.

{}From the present data, we cannot unambiguously determine whether
evolution in the Butcher--Oemler ``blue population'' (mostly disk
galaxies) is related to or independent of the properties of the red
envelope E/S0s, but the evidence for such a connection is not
conspicuous.  A clear difference in the colors of the E/S0 population
between the two clusters would strongly suggest a connection between
the evolution of disk and elliptical galaxies, such as would be
expected if the spiral population merged to form ellipticals.  But the
ellipticals in Abell~851 have only marginally bluer optical--IR colors
than those in Abell~370, and the colors in both clusters are
essentially consistent with simple passive evolution.  There appear to
be greater differences in the IR colors between the two $z \approx 0.4$
clusters, but the poor match to the IR colors of Coma galaxies and to
the model spectra remains unexplained, leaving us unable to draw any
clear inference from this effect.  The lack of evidence in support of
the merger hypothesis is in agreement with the view of Charlot and Silk
(1994) based on their theoretical expectations for z$\sim$0.4 cluster
populations.  Our findings could also be in agreement with the result
of Bothun and Gregg (1990) that the blue B--O galaxies are S0s
undergoing recent star formation at moderate redshifts.

\section{Summary}

Although they are both rich clusters with similar lookback times,
X--ray maps and large--field optical images indicate that Abell 370 is
a more evolved cluster than is Abell 851, in the sense of showing a
more relaxed distribution of galaxies and hot gas.  The greater
maturity appears to be reflected in a more highly evolved population of
disk galaxies in Abell 370.  Our infrared selected samples show that
the disk galaxies in Abell 851 are bluer than those in Abell 370, and
represent a greater fraction of the total sample, suggesting that the
disk galaxies in Abell 370 have had more time to redden and/or fade
into obscurity.  Nevertheless, the early--type galaxies in our samples
lend support to the ``single starburst at high redshift \& passive
evolution'' paradigm for elliptical formation, and do not seem to be
much affected by the differences between the two clusters.  The E/S0s
in both clusters show a clear color--magnitude relation with a slope
and intrinsic scatter similar to that seen in present--epoch Coma
E/S0s, indicating that the early--type galaxies within each cluster
formed at the same time at an early epoch.  The E/S0s in Abell 370 and
851 are slightly bluer in optical$-K$ than those in Coma, by an amount
consistent with passive evolution models for elliptical galaxies formed
at high redshift (BC).  However, the IR colors of an average L$^\ast$
E/S0 in both clusters differ from Coma E/S0s and from elliptical models
in ways that are difficult to explain in terms of age or metallicity
effects.

The analysis described in this paper suggests the ways high quality
optical--IR photometric observations can be used to quantify the
evolutionary histories of cluster ellipticals.  In particular, our data
on clusters at higher redshifts still to be analyzed will be especially
interesting, since spectral evolution is expected to increase the
closer one approaches the era (or eras) of galaxy formation.  Future
papers will address these issues in light of the bulk of our dataset
spanning the redshift range up to $z\sim$ 1.

\acknowledgments
We would like to thank Gus Oemler, Stephane Charlot, Dave Silva, Frank
Valdes, Matt Bershady, and Richard Bower for their assistance through
useful conversations, software support, and sharing data.  This work
was supported in part by the Jet Propulsion Laboratory, California
Institute of Technology, under a contract with the National Aeronautics
and Space Administration.  SAS acknowledges grant support from NAG
5-1607 while at UC-Berkeley. PRME and SAS acknowledge support from a
grant from the NASA IR, Submm, and Radio program, and an $HST$ Cycle 4
GO grant. The authors thank the Aspen Center for Physics for providing
a stimulating environment, as well as desks in a quiet office, in which
some of this paper was written during the Physics of Galaxy Clusters
Workshop in June, 1994.  The authors appreciate the generous support of
NOAO for this project through observing time allocation and assistance.
We are in debt to the Weather for allowing us plenty of time while at
observatories around the world to work on this paper instead of
actually observing.

\begin{figure}

\caption{A plot of the standard rest frame $V,H$ filters, and the observed 
$7520,K$ filters used on Abell 370 in this study.  The observed $K$
band shifts to the rest frame $H$ band, and the observed 7520 band
shifts to rest frame $V$.}

\caption{Comparison of $K$ photometry for the same objects measured on
two different nights for Abell~370.  A theoretical $\pm$1 $\sigma$
errorbar is shown at $K = 16.5$.}

\caption{Comparison of $K$ photometry for the same objects in Abell~370
as measured by AES and in this paper.  The dashed line marks a linear
fit to the differences for $K < 17.5$.}

\caption{Abell~370 $7520,J,K$ composite image, where the colors 
are set to be 7520=blue, J=green, K=red.  The field size is $\sim$4.7
$\times$ 4.7 arcminutes.  The scale in arcsec is shown on the axes with
the zeropoints as defined in Table 2.}

\caption{Abell~851 $7840,J,K$ composite image, where the colors 
are set to be 7840=blue, J=green, K=red.  The field size is $\sim$4.7
$\times$ 4.7 arcminutes.  The scale in arcsec is shown on the axes with
the zeropoints as defined in Table 3.}

\caption{Sample completeness versus $K$ magnitude calculated from
simulated cluster data.}

\caption{Abell~370 (a) and Abell~851 (b), color--magnitude diagrams
for the complete IR samples.  Symbols denote spectroscopic membership
information taken from Soucail et al. (1988) for Abell 370, and from
Dressler and Gunn (1992) for Abell 851: solid squares are members, open
circles are nonmembers, and open triangles do not have redshifts. }
 
\caption{Abell~370 (a) and Abell~851 (b), field--corrected optical--$K$ vs.\
$K$ diagrams, shown together with the appropriately transformed Coma
E/S0 sample.  Typical $\pm 1 \sigma$ errorbars are shown at
representative magnitudes. The $K$ magnitudes in this and all
subsequent c--m diagrams have been corrected by $-0.22$ mag for
seeing.}

\end{figure}

\newpage
\pagebreak[4]

\begin{figure}

\caption{IR/HST samples for Abell~370 (a) and Abell~851 (b) plotted in
optical--$K$ vs.\ $K$ diagrams with morphology and membership indicated
by different symbols.  The open symbols have no spectroscopic
information.  Typical $\pm 1 \sigma$ errorbars are shown at
representative magnitudes.  The best-fit to the B92 Coma E/S0 colors is
shown as the heavy line. }

\caption{SEDs of the average elliptical galaxies brighter than $L^\ast$ in 
Abell~370 and in Abell~851 (after correction for the color-luminosity
relation) relative to the no--evolution and passive--evolution
elliptical models of BC represented by the solid and dashed lines,
respectively.  The average SED of the E/S0s from the Bower Coma sample
are plotted as inverted open triangles.  The vertical error bars on the
cluster points represent $\pm 1 \sigma$ scatter, and the horizontal
bars represent the FWHM of the filters.  The $\pm 1\sigma$ errors on
the median colors are roughly the size of the points for all the
clusters. }

\caption{$H-K$ vs.\ $K$ diagrams for Abell~370 (a) and Abell~851 
(b) IR/HST samples.  Symbols as in Figure 9.  Typical $\pm 1 \sigma$
errorbars are shown at representative magnitudes.  The best-fit to the
B92 Coma E/S0 colors is shown as the heavy line. }

\caption{$J-K$ vs.\ $K$ diagrams for Abell~370 (a) and Abell~851 
(b) IR/HST samples.  Symbols as in Figure 9.  Typical $\pm 1 \sigma$
errorbars are shown at representative magnitudes.  The no--evolution
and passive evolution BC models shown as dashed lines are marked by NE
and PE, respectively.  The models are appropriate only for an L$^\ast$
elliptical; the dashed lines do $not$ indicate a model c--m relation.}

\end{figure}

\end{document}